\documentclass{kybernetika}
\usepackage{amsmath,amsfonts}
\usepackage{algorithmic}
\usepackage{algorithm}
\usepackage{array}
\usepackage[caption=false,font=normalsize,labelfont=sf,textfont=sf]{subfig}
\usepackage{textcomp}
\usepackage{stfloats}
\usepackage{url}
\usepackage{verbatim}
\usepackage{graphicx}
\usepackage{cite}

\newtheorem{theorem}{Theorem}[section]
\newtheorem{lemma}[theorem]{Lemma}

\newtheorem{remark}[theorem]{Remark}

\newtheorem{example}[theorem]{Example}
\newtheorem{definition}[theorem]{Definition}

\newtheorem{assumption}[theorem]{Assumption}
\begin{document}
\pagestyle{myheadings}

\title{Fixed-time safe tracking control of uncertain high-order nonlinear pure-feedback systems via unified transformation functions}

\author{Chaoqun Guo, Jiangping Hu, Jiasheng Hao, Sergej Čelikovský, Xiaoming Hu}


\contact{Chaoqun}{Guo}{School of Automation Engineering, University of Electronic Science and Technology of China, Chengdu 611731, Sichuan, P.\,R. China;  Yangtze Delta Region Institute (Huzhou), University of Electronic Science and Technology of China, Huzhou 313001,  P.\,R. China.}{guochaoqunlg@126.com}
\contact{Jiangping}{Hu\;(Corresponding author)}{School of Automation Engineering, University of Electronic Science and Technology of China, Chengdu 611731, Sichuan, P.\,R. China;  Yangtze Delta Region Institute (Huzhou), University of Electronic Science and Technology of China, Huzhou 313001,  P.\,R. China.}{hujp@uestc.edu.cn}
\contact{Jiasheng}{Hao}{School of Automation Engineering, University of Electronic Science and Technology of China, Chengdu 611731, Sichuan, P.\,R. China.}{hao@uestc.edu.cn}
\contact{Sergej}{Čelikovský}{The Czech Academy 	of Sciences, Institute of Information Theory and Automation, 182 08 Prague 8, Czech Republic}{celikovs@utia.cas.cz}
\contact{Xiaoming}{Hu}{Optimization and Systems Theory, Royal Institute of Technology,
Stockholm SE-10044, Sweden}{hu@kth.se}

\markboth{C. Guo, J. Hu, J. Hao, S. Čelikovský, and X. Hu} {Fixed-time tracking control of systems with state constraints}

\maketitle

\begin{abstract}
	In this paper, a fixed-time safe control problem is investigated for an uncertain high-order nonlinear pure-feedback system with  state constraints. A new nonlinear transformation function is firstly proposed to handle both the constrained and unconstrained cases in a unified way.  Further, a radial basis function neural network is constructed to approximate the unknown dynamics in the system and a fixed-time dynamic surface control (FDSC) technique is developed to facilitate the fixed-time control design for the uncertain high-order pure-feedback system. Combined with the proposed unified transformation function and the FDSC technique, an adaptive fixed-time control strategy is proposed to guarantee the fixed-time tracking. The proposed fixed-time control strategy can guarantee uniform control structure when addressing both constrained and unconstrained situations. Numerical examples are presented to demonstrate the proposed fixed-time tracking control strategy.
\end{abstract}



\section{Introduction}\label{sec:introduction}
Convergence rate has been an important performance index of control systems.  Finite-time controls can ensure that the system state reaches the desired equilibrium in finite time  \cite{H,H2,jjj5} and it has become widely employed  in many practical scenarios, such as variable length pendulum swing \cite{jjj3}, vehicle tracking \cite{jj2,wu21,s3} and finite-time consensus in dynamic networks \cite{jjj4}. However, the so-called {\bf settling time}, needed to reach the equilibrium,  is generally dependent on initial states and no finite bound of settling times is guaranteed for noncompact sets of initial conditions \cite{17,poly12}. To overcome this drawback, the concept of {\it homogeneity in bi-limit} was introduced in \cite{17} to provide conditions for the so-called fixed-time stability, {\it i.e.} the existence of a finite bound of the settling time. Unfortunately, homogeneous approach does not allow to adjust or even estimate the settling time. To overcome this problem, \cite{poly12} introduced a special modification of the so-called
“nested” (terminal) second order sliding mode control algorithm that provided fixed-time stability of the origin and allowed to adjust the
global settling time of the closed-loop system. Motivated by \cite{17,poly12}, the fixed-time stability has been widely studied and applied, see \cite{poly12,s7,s12}  as a sample and  a source of further literature survey.

In this paper, the so-called {\bf pure-feedback systems} will be considered. Pure-feedback systems were introduced and thoroughly studied in \cite{jjj2} along with their more specific version - the so-called {\bf strict feedback system}. Alternatively, the terminology ``triangular form system" was used in the literature, see {\it e.g} \cite{jjj1} and references within there. Note that, \cite{jjj1,jjj2} provided ``classical" asymptotical stabilization only, though \cite{jjj1} used nonsmooth homogeneous approximation, yet with a positive degree, unlike the negative degree homogeneity used for finite-time stability.

Up to now, only a few studies were presented for the fixed-time control of pure-feedback systems having nonaffine connections between cascades, unlike less general strict-feedback systems where the cascade connection is affine and therefore easier to handle. In particular,  \cite{s7,s12} provided the design of the fixed-time controllers for high-order integrator systems and strict-feedback systems only. At the same time, as correctly noted already in \cite{jjj2}, many practical systems are commonly modelled as pure-feedback systems.

\par Besides the fast convergence rate, safety is also a crucial requirement for control systems. In recent years, safe controls have attracted much attention with the development of  practical safety-critical systems, such as robotic systems, chemical plants, and autonomous vehicles \cite{guio,ma}. The output or state variables in safety-critical systems are usually constrained to ensure safeties \cite{jj1}. In order to address safe control problems, barrier Lyapunov function (BLF) methods were commonly applied  in the control design. For example,   log-type BLF \cite{s15}, tan-type BLF \cite{s16} and log-type integral BLF (IBLF) \cite{s17} methods were proposed to deal with static output/full-state constraints. Moreover,  reference \cite{s20}  employed BLF method to tackle a dynamic full-state constraint problem.  However,  BLF-based controls often depend on some feasibility conditions \cite{b1}, which need extra complex offline calculations and even have no solution due to small thresholds of  output/state constraints. In order to overcome such drawbacks, a  nonlinear transformation function (NTF) technique was developed in \cite{s21}, which can transform the original system with state constraints into an unconstrained system. Then, the boundedness of the transformed system can ensure that  the constraints of the original system were satisfied. Moreover, NTF methods do not need additional feasibility conditions.  Therefore, NTF methods have been widely concerned so far.  In  \cite{s22}, a new NTF structure was proposed to solve a tracking control problem for state-constrained strict-feedback systems.

\par It is worthy of noting that in some scenarios, control systems have  constrained and unconstrained states, simultaneously.  Unfortunately, most of the existing safe control strategies are just proposed for control systems with state constraints. Reference \cite{Xujin} introduced a barrier  function to tackle this situation, which relied on some complex feasibility conditions.  Recently, reference  \cite{Song} proposed a nonlinear transformation function to deal with the constrained and unconstrained cases in a unified way. Additionally, another unified nonlinear transformation function was proposed in \cite{Zhang1} for stochastic systems. However, the safety-critical methods presented in the references mentioned above can only achieve asymptotic convergence or finite-time convergence. At the same time, the existing fixed-time controls established for state constrained systems fail in the cases with both  constrained and unconstrained states.   Although reference \cite{Zhang2}  designed a unified fixed-time controller for robotic systems with state constraints, it was only applicable to second-order systems. 

\par As far as we know, fixed-time control of pure-feedback nonlinear systems with both constrained and unconstrained states has not been studied and  suffers from the following two difficulties: construction of nonlinear transformation functions and design of fixed-time controllers. Motivated by the above discussions, this paper attempts to study the fixed-time control of uncertain high-order pure-feedback nonlinear systems with or without state constraints. To overcome the first difficulty, we construct a unified nonlinear transformation function to handle both constrained and unconstrained cases. For the second difficulty, we develop a new fixed-time dynamic surface control (FDSC) technique to facilitate fixed-time control design and reduce computational complexity. 
\par In view of the previous exposition, the contributions of this paper are as follows: 
\par 1) A unified nonlinear transformation function is proposed to transform the original constrained system into an unconstrained one. The proposed nonlinear transformation can ensure that an unconstrained system is a special case of the constrained system. At the same time, the safe fixed-time control problem of the constrained system can be transformed to a fixed-time control problem of an unconstrained system. 
\par 2) A new FDSC technique is developed to facilitate fixed-time control design for the high-order pure-feedback nonlinear system. In contrast to dynamic surface control (DCS) technique proposed in \cite{c1},  the new FDSC technique developed in this paper not only reduces computational cost, but also achieves fixed-time convergence. Compared to the DSC technique proposed in \cite{c2}, the new FDSC technique  can simplify the fixed-time convergence analysis. 
\par 3) Based on the proposed unified nonlinear transformation function and FDSC technique, an adaptive neural network based fixed-time control strategy is proposed for the high-order pure-feedback nonlinear system with unknown dynamics. Thus, the proposed control strategy, in addition to handling the constrained and unconstrained cases in a unified way, even ensures   practical fixed-time tracking for uncertain systems.

\par The rest of the paper is organized as follows. Section \ref{section2} presents some preliminaries and problem formulation.  In Section \ref{section3}, the construction of a unified nonlinear transformation function and the design of an adaptive fixed-time control strategy are given. Simultaneously, practical fixed-time convergence is analyzed for the closed-loop system under the proposed control strategy. Simulation examples are given to validate the proposed fixed-time safe control strategy in Section \ref{section4}. Conclusions are presented in Section \ref{section5}.

\section{Preliminaries and problem formulation}\label{section2}
\subsection{Preliminaries}
Throughout the paper, the following dynamical system having the 
equilibrium at the origin will be considered
\begin{equation}\label{eqj1}
{\dot x}=f(t,x),    f(t,0) \equiv 0,    x\in {\cal D} \subset \mathbb{R}^{n} ,
\end{equation} 
where the right hand side  $f(t,x): \mathbb{R}_{+}\times {\cal D} \mapsto \mathbb{R}^{n}$ 
satisfies the assumptions for the existence of the solution in the 
Fillipov's sense \cite{Filippov}. More specifically, $f(t,x)$ is piecewise 
continuous with respect to $t\ge 0$ for any fixed $x\in {\cal D} $ and 
for any fixed $t>0$ it is continuous with respect to $x\in {\cal D}$ 
except some smooth submanifolds of $ {\cal D}$ where it is 
discontinous and has a finite collection of limit points when $x$ 
approaches that discontinuity manifold.  Here, ${\cal D}$ is a domain 
(open simply connected subset) in $\mathbb{R}^{n}$ containing its origin. Further, 
denote by $x(t,x_{0})$ the solution of (1) such that 
$x(t_{0},x_{0})=x_{0}$, where $t_{0}\ge 0$ is the initial time. Unless 
stated otherwise, in the sequel $t_{0}=0$ and ${\cal D}=\mathbb{R}^{n}.$

\begin{definition}\cite{poly12}
	 The origin of system (1) is said to be globally 
	finite-time stable on $\mathbb{R}^{n}$, if it is globally asymptotically stable 
	and $\forall x_{0} \in \mathbb{R}^{n}$ there exists a positive constant $T(x_{0})$ such that 
	$x(t,x_{0})=0, \forall t\ge T(x_{0})$. The function $T(x_{0}): 
	\mathbb{R}^{n}\mapsto \mathbb{R}_{+}$ is further referred to as the {\bf settling function}.
\end{definition}

\begin{definition}\cite{poly12}
	The origin of system \eqref{eqj1} is said to be fixed-time stable if it is globally finite-time stable and the settling time function $T(x_0)$ is globally 
	bounded on $R^{n}$, {\it i.e.}, $\exists T_{max}>0$: $T(x_0)\leq T_{max}$, $\forall x_0\in\mathbb{R}^n$.
\end{definition}

\begin{definition}
	The origin of system \eqref{eqj1} is said to be practically 
	fixed-time stable on $\mathbb{R}^{n}$, if it is stable and  $\forall \epsilon >0$ 
	there exists a positive constant  $T_{max}(\epsilon )$  such that $\forall x_{0}\in \mathbb{R}^{n} $ 
	there exists $T(x_{0})$, $ T_{max}(\epsilon )> T(x_{0})\ge 0$ and 
	$\forall t > T(x_{0})$ it holds that  $\Vert x(t,x_{0}) \Vert \le 
	\epsilon  $. The function $T(x_{0})$ is further referred as the {\bf 
		practical settling time}.
\end{definition}

\begin{lemma}\cite{Zhang2}\label{lem1}
 The system \eqref{eqj1}  is practically fixed-time stable if $\forall \delta>0$ there exist a  positive definite function $V_{\delta}(t,x)$ and parameters $k_1>0$, $k_2>0$, $0<\gamma<1$, $\beta>1$,  and $0<\theta<1$ such that $$\dot{V}_\delta(t,x)\leq-k_1V_{\delta}(t,x)^\gamma-k_2V_{\delta}(t,x)^\beta+\delta.$$ Furthermore, there exists a settling time $T$ 	such that   $$V_\delta(t,x)\leq\mbox{min}\Big\{\big(\frac{\delta}{k_1\theta}\big)^{\frac{1}\gamma},\big(\frac{\delta}{k_2\theta}\big)^{\frac{1}{\beta}}\Big\},$$
 when $t\geq T$,
 and  the upper bound of the settling time $T$ is given by:
	$$T\leq \frac{1}{k_1(1-\theta)(1-\gamma)}+\frac{1}{k_2(1-\theta)(\beta-1)}.$$

\end{lemma}
\par The following Lemmas are straightforward and were used e.g. in \cite{y1}.
\begin{lemma}\label{lem2}\cite{y1}
	For arbitrary constants $x_1>0$, $x_1\geq x_2$, and $p>1$, it holds:
	$$(x_1-x_2)^p\geq x_2^p-x_1^p.$$
\end{lemma}
\begin{lemma}\label{lem3}\cite{y1}
	For arbitrary constants $p>0$, $x_1\geq 0$, and $x_2>0$, it holds:
	$$x_1^p(x_2-x_1)\leq \frac{1}{1+p}(x_2^{1+p}-x_1^{1+p}).$$
\end{lemma}
\begin{lemma}\label{lem4}\cite{y1}
	For arbitrary constants $x_i\in\mathbb{R}$ and  $p>0$, it holds:
	$$(\sum_{i=1}^{n}|x_i|)^p\leq \mbox{max}(n^{p-1},1)\sum_{i=1}^n|x_i|^p.$$
\end{lemma}

\par Radial basis function neural networks are widely employed to approximate the unknown continuous nonlinear functions in the fields of adaptive control and machine learning. A linearly parameterized model can be used to approximate an unknown continuous function $F(x)\in \mathbb{R}$ as follows:
\begin{equation}\label{eq4}
	F(x)=W^{\top }S(x)+\varepsilon (x), \; x\in \mathbb{R}^{n}, 
\end{equation}
where $W\in \mathbb{R}^{N}$ is the weight vector of 
a radial basis function neural network and   
$S(x)=[S_{1}(x),...,S_{N}(x)]^T\in\mathbb{R}^N$ is the basis function vector. More specifically, 
\begin{equation}
	S_{i}(x)=\mbox{exp}\big[-\frac{(x-\tau_{i})^T(x-\tau_{i})}{\psi_i^2}\big],\quad i=1,...,N,
\end{equation}
where $\psi_i\in\mathbb{R}$, $\tau_{i}\in\mathbb{R}^n$ are the so-called width and the 
so-called center of the basis function, respectively. Finally,  $\varepsilon\in\mathbb{R}$ is the estimation error. 
\begin{assumption}\label{ass4}
	In the linearly parameterized model \eqref{eq4}, $\|W\|\leq \bar{W}$, $|\varepsilon|\leq \varepsilon_{1}$,  where $\bar{W}$ and $\varepsilon_{1}$ are unknown positive constants. 
	In the sequel, denote  $w=\mbox{max}\{\bar{W},\varepsilon_{1}\}$.
\end{assumption}

\subsection{Problem formulation}
Consider the following pure-feedback 
system \cite{jjj1}: 
\begin{equation}\label{eq1}
	\begin{cases}
		\dot{x}_i=f_i(\bar{x}_i,x_{i+1}), i=1,...,n-1\\
		\dot{x}_n=f_n(\bar{x}_n,u),\\
		y=x_1,
	\end{cases}
\end{equation}
where $x=[x_1,...,x_n]^T\in\mathbb{R}^n$ is the state; $\bar{x}_i$ denotes $[x_1,...,x_i]^T\in\mathbb{R}^i$; $y\in\mathbb{R}$ is the scalar output;  $u\in\mathbb{R}$ is the the scalar control input and $f_i(\cdot)$\; $(i=1,...,n)$ are unknown continuous nonlinear  functions. 
The system is required to satisfy 
the following state constraints:
\begin{equation}\label{eq2}
	-h_{i1}(t)<x_i(t)<h_{i2}(t), \;i=1,...,n,
\end{equation}
where the time-varying bounds $h_{i1}(t)$ and $h_{i2}(t)$ are strictly positive functions.

\begin{assumption}
	The functions $h_{i1}(t), h_{i2}(t)$ and their 
	derivatives are uniformly bounded on $\mathbb{R}_{+}$.
\end{assumption}
\begin{assumption}\label{ass5}
	The initial states satisfy $-h_{i1}(0)<x_i(0)<h_{i2}(0)$ for $i=1,...,n$.
\end{assumption}
\begin{assumption}\label{ass1}
	$f_i(\bar{x}_i,x_{i+1})\;(i=1,...,n-1)$ and $f_n(\bar{x}_n,u)$ are continuously differentiable for all $ x\in \mathbb{R}^{n}$. 
\end{assumption}

\begin{assumption} \label{ass2}
	Let  $g_{n}(\bar{x}_{n},u):= \frac{\partial f_{n}(\bar{x},u)}{\partial u}$. It holds that $\underline{g}_n\leq g_n(\cdot)\leq\overline{g}_n$, where $\underline{g}_n$, $\overline{g}_n$ are unknown positive constants.
\end{assumption}

Further, 
assume that the reference output $y_{d}$ is given to be followed by  the 
output of the nonlinear  system (4).
\begin{assumption}\label{ass3}
	The reference output $y_d(t)$ satisfies the constraint (\ref{eq2}), i.e., $-h_{11}(t)<y_d(t)<h_{12}(t)$.  Moreover, the reference output $y_{d}(t)$, its first and its second  
	derivatives are uniformly bounded on $\mathbb{R}_{+}$.
\end{assumption}
\par The aim of this paper is to design a practical 
fixed-time controller for the system (\ref{eq1}) providing the given reference output tracking, that is,
$$|y(t)-y_d(t)|\leq \zeta, \quad \forall t>T,$$ where $T, \zeta $ are some positive constants and at the same 
time the designed controller guarantees that the state constraints \eqref{eq2}
are not violated at any time.

\section{Main results}\label{section3}
\par This section is divided into two parts. Subsection \ref{3.1} introduces a unified nonlinear transformation function to transform the original constrained system \eqref{eq1} to a new unconstrained system. Subsection \ref{3.2} gives the practical fixed-time controller to achieve the 
previously formulated tracking goal.

\subsection{Unified nonlinear transformation function}\label{3.1}
Consider the system \eqref{eq1} and the constraint \eqref{eq2}. A unified nonlinear transformation function is proposed as follows:
\begin{equation}\label{eq6}
	\xi_{i}(t)= \frac{h_{i1}(t)+h_{i2}(t)}{4} \ln \frac{h_{i1}(t)+x_{i}(t)}{h_{i2}(t)-x_{i}(t)}.
\end{equation}
The nonlinear transformation  \eqref{eq6} 
has the following property. For any $t\ge 0$ it holds:
\begin{equation}
	\begin{cases}
		(\mbox{\romannumeral1}) \lim\limits_{x_i(t)\to -h_{i1}(t)}\xi_i(t)=-\infty;\\
		(\mbox{\romannumeral2}) \lim\limits_{x_i(t)\to h_{i2}(t)}\xi_i(t)=+\infty;\\
		(\mbox{\romannumeral3}) \lim\limits_{h_{i1}(t)=h_{i2}(t)\to+\infty}\xi_i(t)=x_i(t).
	\end{cases}
\end{equation}
 The properties (\romannumeral1) and (\romannumeral2) can be easily obtained. When $h_{i1}(t)=h_{i2}(t)\to+\infty$, by 
straightforward computations, we can also verify that the property (\romannumeral3) holds.

\par According to the properties (1) and (2), one can obtain that if $\xi_i(t)$ is bounded, then the condition $-h_{i1}(t)<x_i(t)<h_{i2}(t)$ holds for any $-h_{i1}(0)<x_i(0)<h_{i2}(0)$. Therefore,  in order to ensure that the constraints are not violated, just ensure that $\xi_i(t)$ is bounded. For the property (3), $h_{i1}(t)=h_{i2}(t)=+\infty$ means that there is no state constraint and thus $\xi_i(t)=x_i(t)$. Consequently, the proposed  transformation \eqref{eq6} can deal with the cases with and without state constraints in a unified manner.
\begin{remark}
	The transformation function is different from the existing transformation functions \cite{s21,s22}, which can only deal with the case with state constraints. Actually, the proposed transformation function \eqref{eq6} can be applied to systems in the following situations: 1) All states are constrained; 2) All states are not constrained; 3) Some states are constrained and the  others are not constrained. Therefore, the proposed NTF \eqref{eq6} can deal with the constrained and unconstrained states in a unified way and facilitate the design of a unified structured controller.
\end{remark}

Using \eqref{eq1}, \eqref{eq6},  and differentiating $\xi=[\xi_1,...,\xi_n]^T$ yields:
\begin{equation}\label{eq9}
	\begin{cases}
		\dot{\xi}_i=\varphi_i[F_i(\bar{x}_{i+1},\xi_{i+1})+\xi_{i+1}]+\psi_{i},\,\,i=1,...,n-1,\\
		\dot{\xi}_n=\varphi_n[F_n(\bar{x}_n)+g_nu]+\psi_n,
	\end{cases}
\end{equation}
where 
\begin{equation}\nonumber
	\varphi_i=\frac{\partial{\xi_i}}{\partial{x_i}}=\frac{\big(h_{i1}(t)+h_{i2}(t)\big)^2}{4\big(h_{i1}(t)+x_i(t)\big)\big(h_{i2}(t)-x_i(t)\big)},
\end{equation}
\begin{equation}\nonumber
	\begin{aligned}
	\psi_i=&\frac{h_{i1}(t)+h_{i2}(t)}{4}\Big(\frac{\dot{h}_{i1}(t)}{h_{i1}(t)+x_i(t)}-\frac{\dot{h}_{i2}(t)}{h_{i2}(t)-x_i(t)}\Big)+\frac{\dot{h}_{i1}(t)+\dot{h}_{i2}(t)}{4}\ln \frac{h_{i1}(t)+x_{i}(t)}{h_{i2}(t)-x_{i}(t)},
	\end{aligned}
\end{equation}
\begin{equation}\nonumber
	F_{i}(\bar{x}_{i+1},\xi_{i+1})=f_i(\bar{x}_i,x_{i+1})-\xi_{i+1},\,\,i=1,...,n-1,
\end{equation}
$$F_n(\bar{x}_n)=f_n(\bar{x}_n),$$
and $f_n(\bar{x}_n)=f_n(\bar{x}_n,0)$.
\par Obviously, the original system \eqref{eq1} with the state constraint \eqref{eq2} is transformed  to an unconstrained system \eqref{eq9}. The constraints on the state $x$ can be guaranteed by ensuring the boundedness of the variable $\xi$. In sequel, based on the transformed system \eqref{eq9}, we need to design a  fixed-time controller to not only  ensure that the state constraints are not violated, but also  realize the practical fixed-time output tracking. 
\subsection{Control design and convergence analysis}\label{3.2}
According to the unified nonlinear transformation function \eqref{eq6} and  Assumption \ref{ass3}, a nonlinear transformation function is given for the reference output $y_d$ as follows:
\begin{equation}\label{eq7}
		\xi_{d}(t)= \frac{h_{i1}(t)+h_{i2}(t)}{4}[ \ln (h_{i1}(t)+y_{d})- 
	\ln(h_{i2}(t)-y_{d})].
\end{equation}
\par Define an error system as follows:
\begin{equation}\label{eqq12}
	\begin{cases}
		\zeta_1=\xi_1-\xi_d,\\
		\zeta_i=\xi_i-\alpha_{if}, &i=2,...,n,
	\end{cases}
\end{equation}
where $\alpha_{if}$ is a  dynamic variable designed by the following 
\textbf{fixed-time dynamic  surface control (FDSC) technique}:
\begin{equation}\label{eq10}
	\lambda_i\dot{\alpha}_{if}=(\alpha_{i-1}-\alpha_{if})^{r_1}+(\alpha_{i-1}-\alpha_{if})^{r_2}+\alpha_{i-1}-\alpha_{if},\quad i=2,...,n,
\end{equation}
where $r_1=\frac{m}{n}$, $r_2=\frac{p}{q}$, and  $m<n$, $p>q$ are positive odd constants, $\lambda_{i}$ is a positive constant, and
$\alpha_{i-1}$, $i=2,...,n$, are the virtual controllers to be designed latter.	
\begin{remark}
	Obviously, the equilibrium of \eqref{eq10} is $\alpha_{i-1}$, this means that the $\alpha_{if}$ converges to the $\alpha_{i-1}$ at the steady state. Thus, $\alpha_{if}$ in \eqref{eq10} can be regarded as an estimation of  $\alpha_{i-1}$, and its derivative $\dot{\alpha}_{if}$ is an estimation of $\dot{\alpha}_{i-1}$. Therefore, the FDSC technique \eqref{eq10} is used to generate the derivative  $\dot{\alpha}_{if}$ to replace $\dot{\alpha}_{i-1}$ appearing in the backstepping design process of the controller, so there is no need to calculate the derivative of the virtual controller $\alpha_{i-1}$, which can reduce the computational complexity.  In contrast to the DSC   in \cite{c1},  the  FDSC \eqref{eq10} not only reduces computational cost, but also has the characteristic of fixed-time convergence. 	
	Moreover, compared with the DSC technique in \cite{c2}, the FDSC technique \eqref{eq10} has additional term $\alpha_{i-1}-\alpha_{if}$, which obviously leads to faster estimation speed and facilitate to the fixed-time convergence of the system. 
\end{remark}

\par According to the error system \eqref{eqq12} and backstepping-like design method, we design the controller $u$ as follows: 
\begin{equation}\label{eqq41}
	\begin{split}
		u=&-\rho_n\varphi_n\hat{a}_n\mu_n^2\zeta_n-\frac{\psi_n^2\zeta_n}{\varphi_n}-\frac{\dot{\alpha}_{nf}^2\zeta_n}{\varphi_n}\\&-\frac{\varphi_{n-1}^2\zeta_{n-1}^2\zeta_n}{\varphi_n}-\frac{k_{n1}}{\varphi_n}\zeta_n^{r_1}-\frac{k_{n2}}{\varphi_n}\zeta_n^{r_2},
	\end{split}
\end{equation}

\begin{equation}\label{eqq422}
	\dot{\hat{a}}_n=\rho_n\varphi_n^2\mu_n^2\zeta_n^2-\sigma_{n1}\hat{a}_n^{r_1}-\sigma_{n2}\hat{a}_n^{r_2},
\end{equation}
where $\rho_n$, $\sigma_{n1}$, and  $\sigma_{n2}$ are positive constants, $\mu_n$ is a positive variable defined in controller design process.  The design process of the controller $u$ is shown in Appendix \ref{appendix1}.

\par Under the controller \eqref{eqq41}, 
the main control results are shown in Theorem \ref{th1}

\begin{theorem}\label{th1}
	Consider the uncertain high-order pure-feedback system \eqref{eq1} with the state constraints given by \eqref{eq2}. Under the Assumptions \ref{ass4}- \ref{ass1}, the proposed controller \eqref{eqq41} can achieve the following objectives:
	\par (1) The closed-loop system is practically fixed-time stable.
	\par (2) The output tracking error $e=y-y_d$ is bounded in fixed time.
	\par (3) The state constraints are satisfied all the time.
\end{theorem}
Proof. See the Appendix \ref{appendix2}.

\section{Simulation example}\label{section4}
In this section, two examples are presented to demonstrate the proposed fixed-time safe control strategy. Example \ref{example1} verifies the effectiveness of the fixed-time control strategy for the system with full state constraints. Example \ref{example2} shows that the proposed control strategy can still work for the system with partial state constraints. 
\begin{example}\label{example1}
	Consider the following pure-feedback nonlinear system:
	\begin{equation}\label{eq50}
		\begin{cases}
			\dot{x}_1=x_1+x_2+x_2^2,\\
			\dot{x}_2=x_1x_2+u+0.1\sin(u),\\
			y=x_1.
		\end{cases}
	\end{equation}

The state constraint functions are  given by $h_{11}(t)=0.5+0.3\sin(t)$, $h_{12}(t)=0.6-0.2\sin(t)$, $h_{21}(t)=0.5+0.2\cos(t)$, and  $h_{22}(t)=0.5+0.2\cos(t)$. The reference output is  given by $y_d=0.1\sin(0.5t)$.   
\par The initial states are given as $x(0)=[0.2,-0.2]^T$. The parameters in  the controller \eqref{eq41} are selected as follows: $k_{11}=k_{12}=k_{21}=k_{22}=2$, $\sigma_{11}=0.2$, $\sigma_{12}=0.5$, $\sigma_{21}=0.3$, $\sigma_{22}=0.4$, $\rho_1=\rho_2=0.0005$, $\lambda_2=0.1$, $r_{1}=97/99$, and  $r_2=99/97$. The parameters of the neural network approximation of $F_1(X_1)$ are choosen as follows:
$\psi_i=3$, $\tau_i=[-\tau_{ii},0,\tau_{ii}]^T\; (i=1,...,5)$. The parameters of the  neural network approximation of $F_2(X_2)$ are choosen as follows: $\psi_i=3$, $\tau_i=[-\tau_{ii},\tau_{ii}]^T\;(i=1,...,5)$. Here, $\tau_{ii}$ are selected as $1,2,3,4$ and $5$, respectively. \begin{figure}[!htbp]
	\centering
	\includegraphics[height=4.8cm]{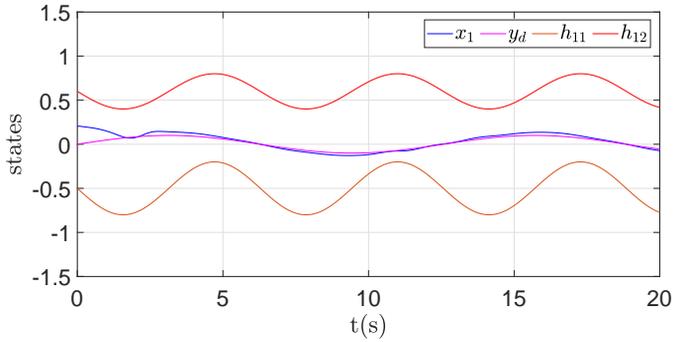}
	\caption{The trajectories of the state $x_1$ and the reference output $y_d$.}
	\label{figure1}
\end{figure}
\begin{figure}[!htbp]
	\centering
	\includegraphics[height=4.8cm]{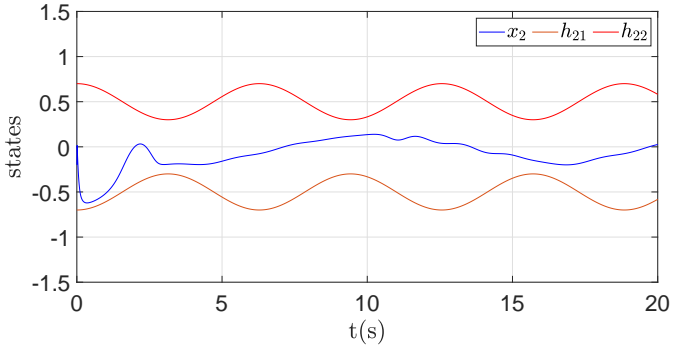}
	\caption{The trajectory of the state $x_2$.}
	\label{figure2}
\end{figure}

\par Fig. \ref{figure1} illustrates the trajectories of the state $x_1$ and the desired output $y_d$. In Fig. \ref{figure1}, we can see that state $x_1$ can track the reference output $y_d$ at about $T=5s$ under the proposed safe controller (\ref{eq41}). Moreover, the state $x_1$ satisfies the given constraint all the time. 
Fig. \ref{figure2} illustrates the trajectory of the state $x_2$. From Fig. \ref{figure2}, we can see that $x_2$ is always within the constraint. Fig. \ref{figure3} shows the trajectory of the tracking error and demonstrates that the output tracking error is bounded in fixed time, that is, $|e|\leq 0.05$ for $t\geq5s$. Moreover, the control input $u$ is shown in Fig. \ref{figure_u1}.

\begin{figure}[!htbp]
	\centering
	\includegraphics[height=4.8cm]{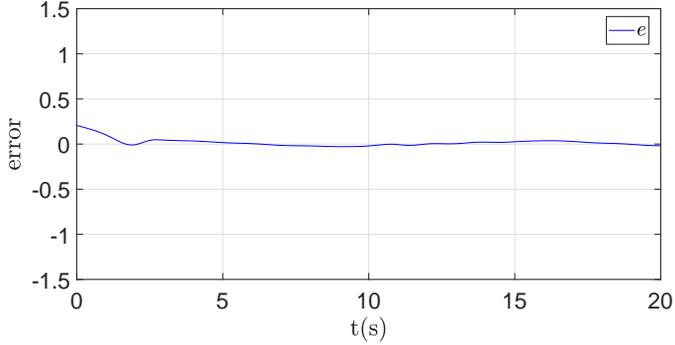}
	\caption{The trajectory of the output tracking error $e$.}
	\label{figure3}
\end{figure}

\begin{figure}[!htbp]
	\centering
	\includegraphics[height=4.8cm]{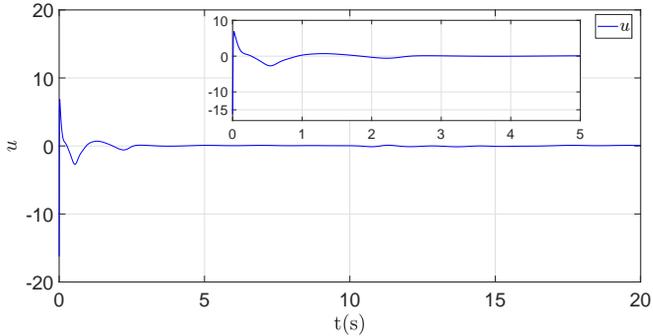}
	\caption{The trajectory of controller $u$.}
	\label{figure_u1}
\end{figure}

\end{example}

\begin{figure}[!htbp]
	\centering
	\includegraphics[height=4.8cm]{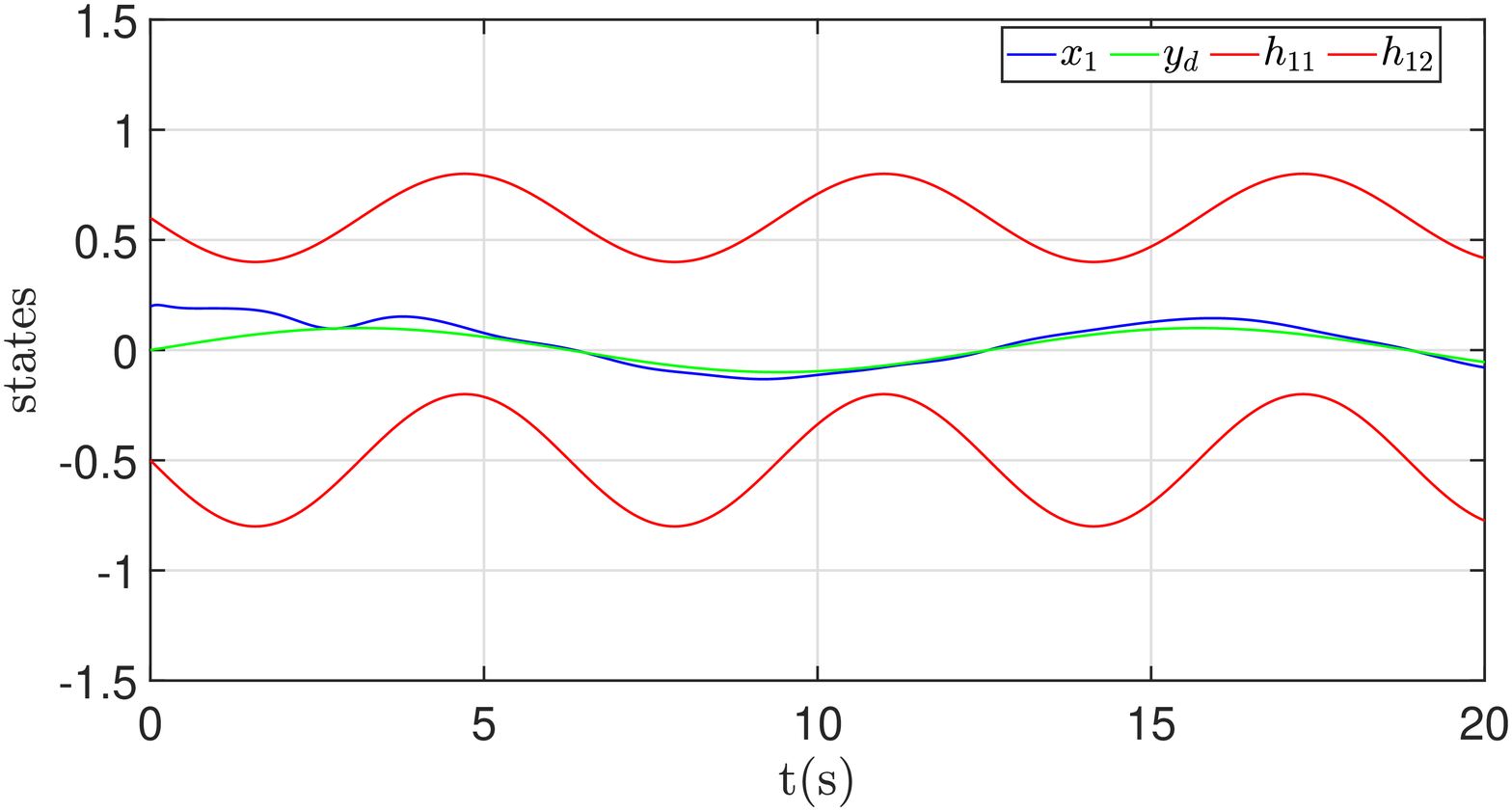}
	\caption{The trajectories of the state $x_1$ and the  reference output $y_d$.}
	\label{figure5}
\end{figure}

\begin{figure}[!htbp]
	\centering
	\includegraphics[height=4.8cm]{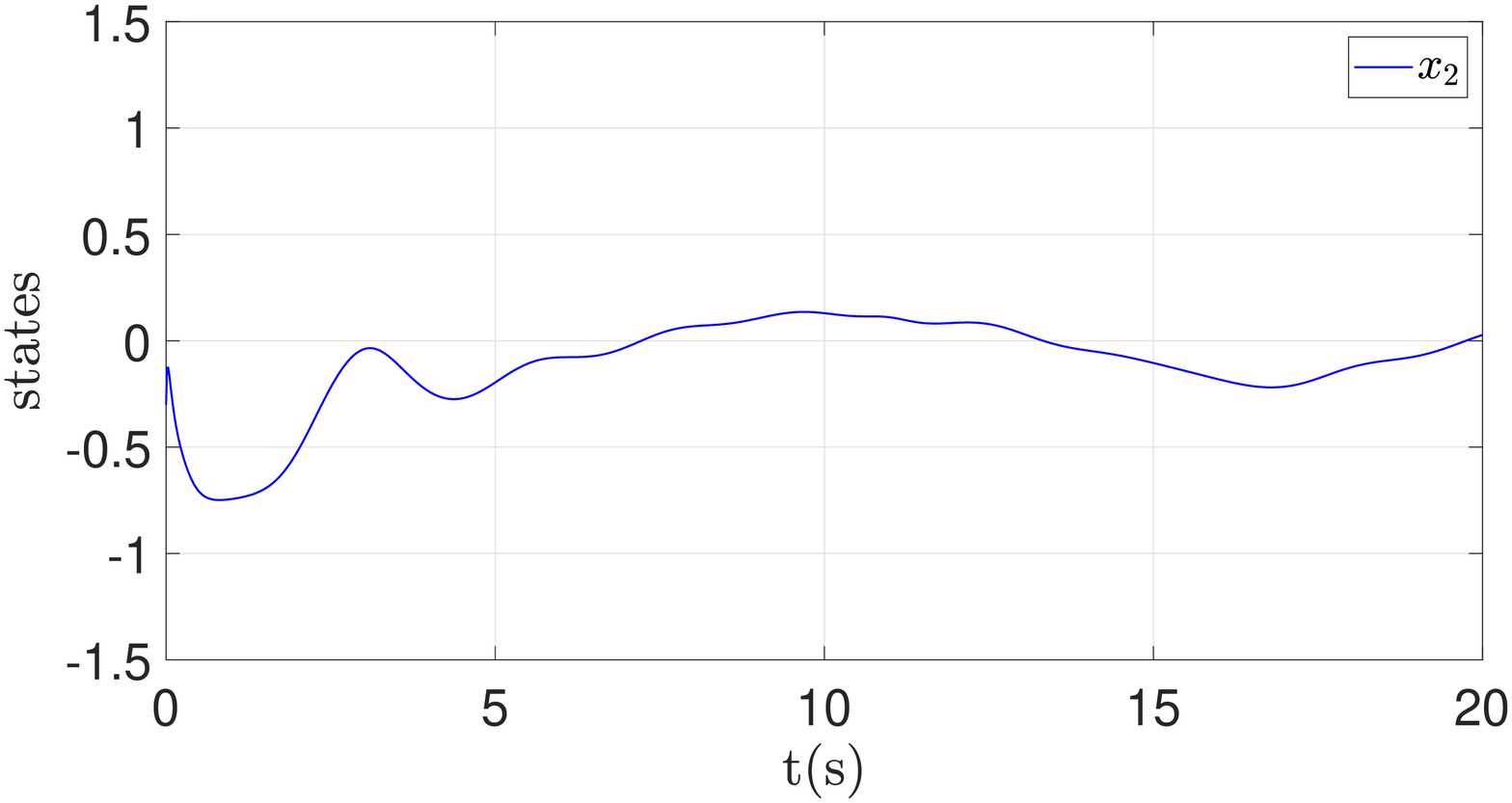}
	\caption{The trajectory of the state $x_2$.}
	\label{figure6}
\end{figure}

\begin{figure}[!htbp]
	\centering
	\includegraphics[height=4.8cm]{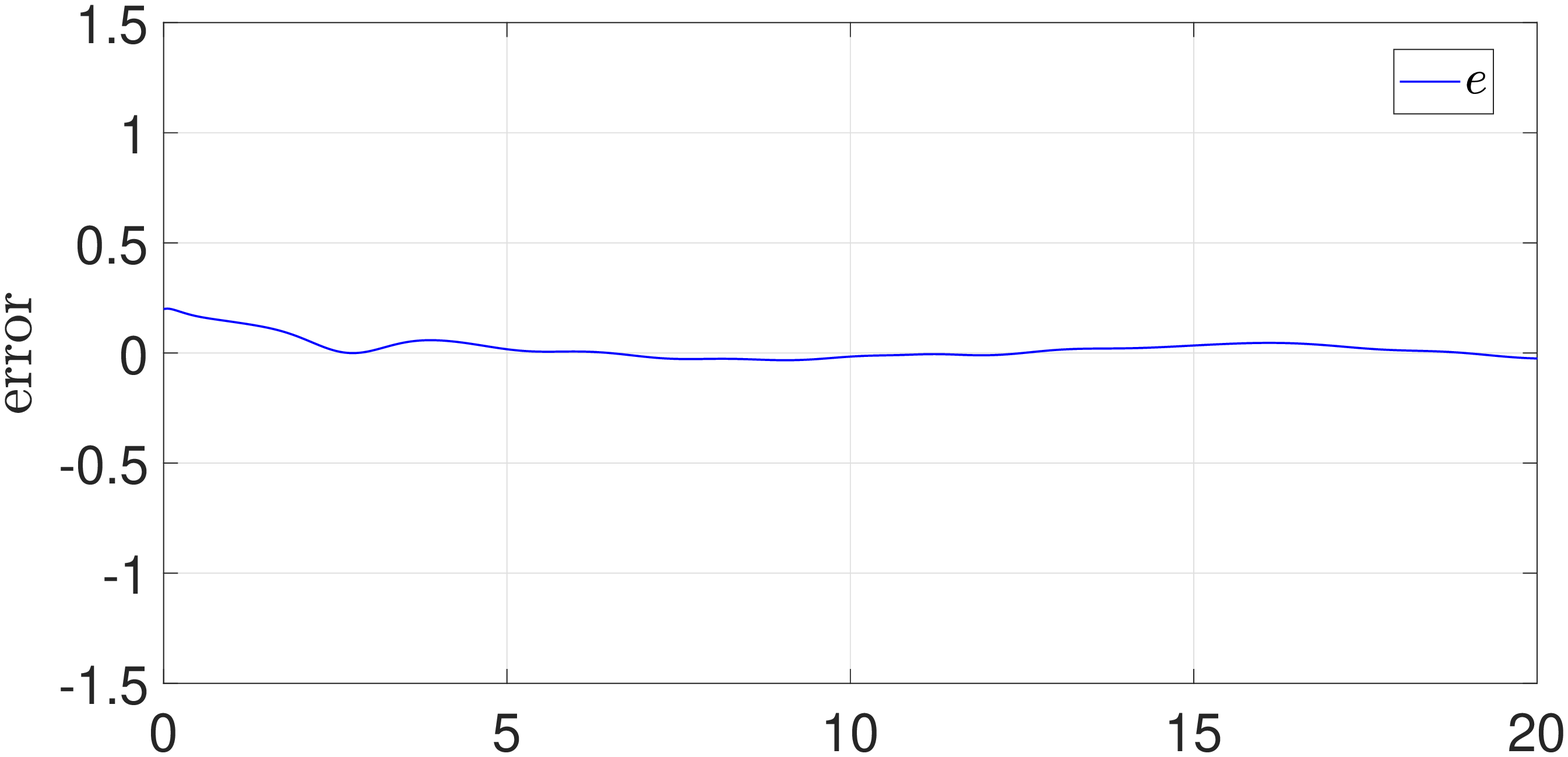}
	\caption{The trajectory of the output racking error $e$.}
	\label{figure7}
\end{figure}
\begin{figure}[!htbp]
	\centering
	\includegraphics[height=4.8cm]{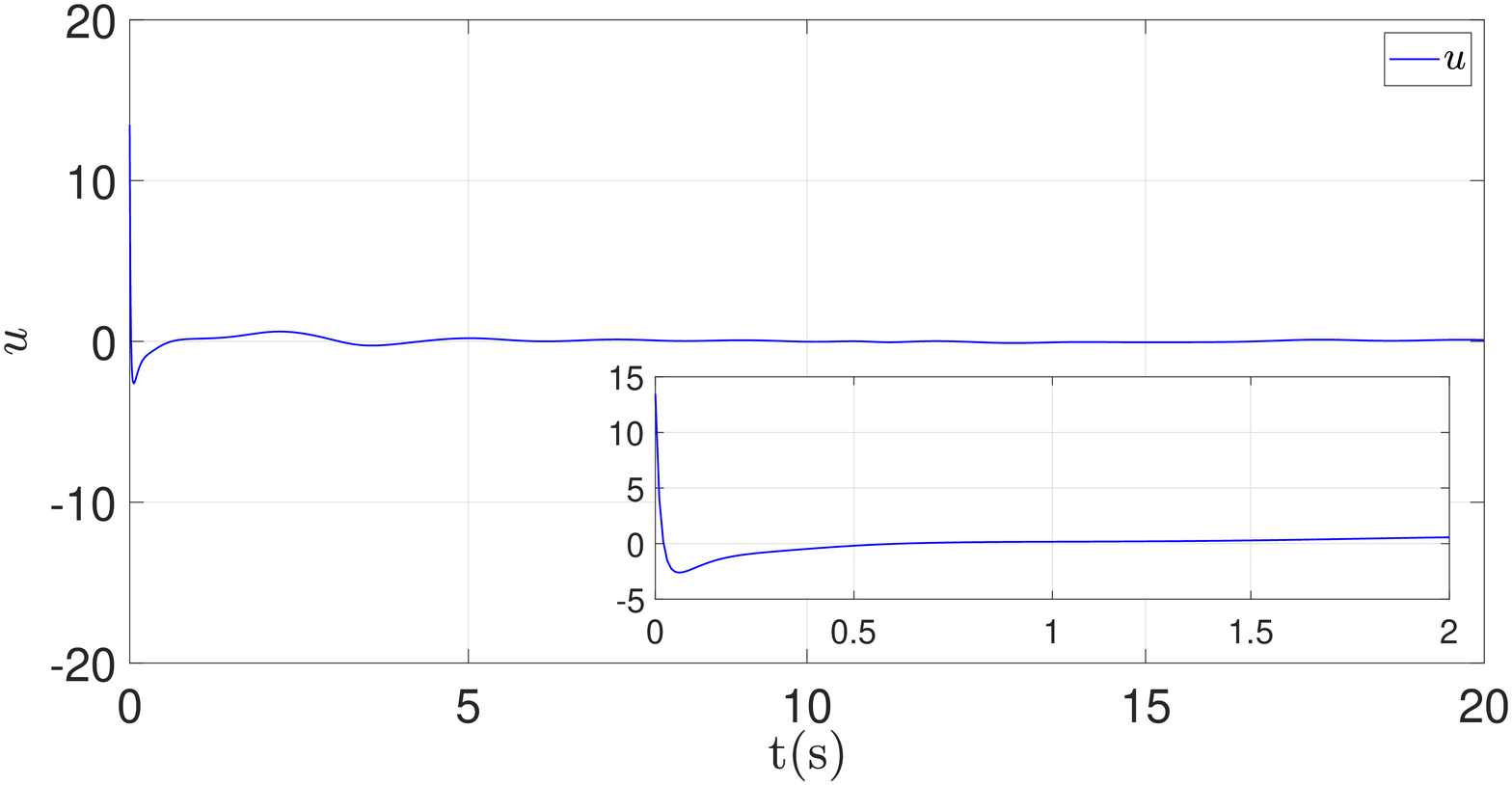}
	\caption{The trajectory of the control input $u$.}
	\label{figure_u2}
\end{figure}
\begin{example}\label{example2}
For the pure-feedback nonlinear system \eqref{eq50}, we assume that the state $x_1$ is constrained by $h_{11}(t)=0.5+0.3\sin(t)$ and $h_{12}(t)=0.6-0.2\sin(t)$, while the state $x_2$ has no constraint, that is $h_{21}(t)=h_{22}(t)\equiv+\infty$. The initial state is given by $x(0)=[0.2,-0.3]^T$. Set $k_{11}=2$, $k_{12}=1$, $k_{21}=k_{22}=2$. Other parameters are similar with those in Example \ref{example1}.

The trajectories of the states $x_1$ and $x_2$ are shown in Fig. \ref{figure5} and Fig. \ref{figure6}, respectively. From  Fig. \ref{figure5} and Fig. \ref{figure6}, it can be found that  $x_1$ and $x_2$ are bounded while $x_1$ does not violate its constraint all the time. Moreover, the output tracking error $|e|\leq 0.05$ for $t\geq 4.3s$, which is illustrated in Fig. \ref{figure7}. Fig. \ref{figure_u2} shows the response of the controller $u$.
\par
The above simulation results  show that the proposed fixed-time safe controller can deal with the output tracking problem for the uncertain pure-feedback nonlinear system with and without state constraints while keeping one control structure.
\end{example}

\section{Conclusions}\label{section5}
This paper has solved the fixed-time output tracking problem for uncertain high-order pure-feedback systems with and without state constraints in a unified way. A nonlinear transformation function method has been proposed to deal with the cases with and without state constraints. With the help of the unified nonlinear transformation, the fixed-time safe control problem has been transformed to just  a fixed-time control problem. At the same time, a fixed-time dynamic surface control technique has been developed to facilitate the fixed-time controller design. Thus, an adaptive neural network based fixed-time control strategy has been proposed for the uncertain pure-feedback nonlinear system. Theoretical results have shown that the fixed-time convergence of the output tracking error can be achieved and all the state constraints can always be satisfied under the proposed control strategy. Moreover, the effectiveness of the proposed control strategy has been validated by numerical simulations.

\section*{Acknowledgment}
This work was partially supported by the Czech Science Foundation under Grant No. 21-03689S,  National Natural Science Foundation of China under Grants No.62103341, No. 61473061, No. 71503206, No. 61104104, and the Sichuan Science and Technology Program under Grant 2020YFSY0012.

\section*{Appendix}
\appendix

\section{Design process of the controller $u$}\label{appendix1}
In this subsection, we present the design process of the controller $u$ by \textbf{backstepping-like design method}. First, define 

\begin{equation}\label{eq12}
	y_i=\alpha_{if}-\alpha_{i-1},\quad i=2,...,n,
\end{equation}
 where $\alpha_{if}$ is generated by the \textbf{FDSC technique} in \eqref{eq10}, $\alpha_{i-1}$ is the virtual controller designed latter. Next, we design the virtual controller $\alpha_{i-1}$ at the $(i-1)$th step, and the controller $u$ at the $n$th step.
\par \textbf{Step 1:} Differentiating $\zeta_1$ yields: 
\begin{equation}
	\dot{\zeta}_1=\varphi_1[F_1(x_1,x_2,\xi_2)+\xi_2]+\psi_1-\dot{\xi}_d.
\end{equation}
According to \eqref{eqq12} and \eqref{eq12}, it follows that 
$\xi_2=\zeta_2+y_2+\alpha_{1}$. Then, we have
\begin{equation}
	\dot{\zeta}_1=\varphi_1[F_1(x_1,x_2,\xi_2)+\zeta_2+y_2+\alpha_{1}]+\psi_1-\dot{\xi}_d.
\end{equation}
\par Using the neural network based approximation \eqref{eq4},  $F_1(x_1,x_2,\xi_2)=F_1(X_1)$ can be approximated as follows:
\begin{equation}
	F_1(X_1)=W_1^TS(X_1)+\varepsilon_1(X_1),
\end{equation}
where $X_1=[x_1,x_2,\xi_2]^T$, $W_1\in\mathbb{R}^N$ and $S(X_1)\in\mathbb{R}^N$ are the ideal weight vector and basis function vector respectively, and 
$\varepsilon_1(X_1)$ is the estimation error. According to Assumption \ref{ass4}, we have $\vert|W_1\vert|\leq \bar{W}_1$, $|\varepsilon_1|\leq\varepsilon_{11}$, where $\bar{W}_1$ and $\varepsilon_{11}$ are unknown positive constants. 
\par Let $w_1=\mbox{max}\{\bar{W}_1,\varepsilon_{11}\}$, then one has $$F_1(X_1)\leq w_1\mu_1(X_1),$$
where $\mu_1(X_1)=\vert|S(X_1)\vert|+1$.
\par Next,  the virtual controller $\alpha_{1}$ is designed as follows:
\begin{equation}\label{eq16}
	\begin{split}
		\alpha_1=&-\rho_1\hat{a}_1\varphi_1\mu_1^2\zeta_1-\varphi_1\zeta_1+\frac{\dot{\xi}_d}{\varphi_1}-\frac{\psi_1}{\varphi_1}\\&-\frac{k_{11}}{\varphi_1}\zeta_1^{r_1}-\frac{k_{12}}{\varphi_1}\zeta_1^{r_2},
	\end{split}
\end{equation}
where $k_{11}$, $k_{12}$, and  $\rho_1$ are  positive constants,  $r_1$ and $r_2$ are defined in \eqref{eq10}. Let $\hat{a}_1$ be the estimation of $a_1=w_1^2$, which is determined by the following adaptive law:
\begin{equation}\label{eq17}
	\dot{\hat{a}}_1=\rho_1\varphi_1^2\mu_1^2\zeta_1^2-\sigma_{11}\hat{a}_1^{r_1}-\sigma_{12}\hat{a}_1^{r_2},
\end{equation}
where $\sigma_{11}$ and $\sigma_{12}$ are positive constants.
\par According to Young's inequality, we have the following inequalities:
\begin{equation}\label{eq19}
	\begin{split}
		\varphi_1\zeta_1F_1\leq|\varphi_1\zeta_1w_1\mu_1|\leq \rho_1w_1^2\varphi_1^2\mu_1^2\zeta_1^2+\frac{1}{4\rho_1},
	\end{split}
\end{equation}
\begin{equation}\label{eq20}
	\varphi_1\zeta_1y_2\leq\varphi_1^2\zeta_1^2+\frac{y_2^2}{4},
\end{equation}
where $\rho_1$ is defined in \eqref{eq16}.
\par Then, we construct a Lyapunov function as follows:
\begin{equation} V_1=\frac{1}{2}\zeta_1^2+\frac{1}{2}\tilde{a}_1^2+\frac{1}{2}y_2^2,
\end{equation}
where $\tilde{a}_1=a_1-\hat{a}_1$. 

Calculating the derivative of $V_1$  and employing the virtual controller \eqref{eq16}, and  inequalities \eqref{eq19}, \eqref{eq20}, we have
\begin{equation}\label{eq21}
	\begin{split}
		\dot{V}_1\leq&\rho_1\tilde{a}_1\varphi_1^2\mu_1^2\zeta_1^2+\varphi_1\zeta_1\zeta_2-k_{11}\zeta_1^{1+r_1}-k_{12}\zeta_1^{1+r_2}+\frac{y_2^2}{4}\\&
		+\frac{1}{4\rho_1}+\tilde{a}_1\dot{\tilde{a}}_1+y_2\dot{y}_2\\
		=&\rho_1\tilde{a}_1\varphi_1^2\mu_1^2\zeta_1^2+\varphi_1\zeta_1\zeta_2-k_{11}\zeta_1^{1+r_1}-k_{12}\zeta_1^{1+r_2}+\frac{y_2^2}{4}\\&
		+\frac{1}{4\rho_1}-\tilde{a}_1\dot{\hat{a}}_1+y_2\dot{y}_2.
	\end{split}
\end{equation}
\par Using the adaptive  law \eqref{eq17}, we have
\begin{equation}\nonumber
	\begin{split}
		-\tilde{a}_1\dot{\hat{a}}_1=-\rho_1\tilde{a}_1\varphi_1^2\mu_1^2\zeta_1^2+\sigma_{11}\tilde{a}_1\hat{a}_1^{r_1}+\sigma_{12}\tilde{a}_1\hat{a}_1^{r_2}.
	\end{split}
\end{equation}
Since $\tilde{a}_1=a_1-\hat{a}_1$, thus $\sigma_{11}\tilde{a}_1\hat{a}_1^{r_1}$ and $\sigma_{12}\tilde{a}_1\hat{a}_1^{r_2}$ can be written as $\sigma_{11}\hat{a}^{r_1}(a_1-\hat{a}_1)$ and $\sigma_{12}\hat{a}_1^{r_2}(a_1-\hat{a}_1)$, respectively. By analyzing the adaptive  law \eqref{eq17}, it can be verified that for any given initial value $\hat{a}_1(0)\geq0$, one has  $\hat{a}_1(t)\geq0$ if $\dot{\hat{a}}_1(t)\geq0$. It is noted that  $\sigma_{11}$ and $\sigma_{12}$ are positive constants. If $\dot{\hat{a}}_1(t)<0$, then $\hat{a}_1(t)$ will decrease until ${\hat{a}}_1(t)=0$ at a certain time $t_d$. Due to the fact that $\rho_1\varphi_1^2\mu_1^2\zeta_1^2\geq0$, it can be found that $\dot{\hat{a}}_1(t)\geq0$ when $\hat{a}_1(t)=0$. Therefore, $\hat{a}_1(t)\geq0$ after $t\geq t_d$. Thus, if we choose an initial value $\hat{a}_1(0)\geq0$, then we have $\hat{a}_1(t)\geq0$.
\par According to  Lemma \ref{lem3}, we  have
\begin{equation}\nonumber
	\begin{split}
		\sigma_{11}\hat{a}_1^{r_1}\tilde{a}_1&=\sigma_{11}\hat{a}_1^{r_1}(a_1-\hat{a}_1)\\&\leq\sigma_{11}\frac{1}{1+r_1}(a_1^{1+r_1}-\hat{a}_1^{1+r_1})\\&=\sigma_{11}\frac{1}{1+r_1}(a_1^{1+r_1}-[a_1-\tilde{a}_1]^{1+r_1}).
	\end{split}
\end{equation}
\par When $\hat{a}_1(t)>0$, we have $a_1(t)>\tilde{a}_1(t)$. From Lemma \ref{lem2}, we have
\begin{equation}
	\begin{split}
		\sigma_{11}\hat{a}_1^{r_1}\tilde{a}_1&\leq \sigma_{11}\frac{1}{1+r_1}(a_1^{1+r_1}+a_1^{1+r_1}-\tilde{a}_1^{1+r_1})\\&=\frac{2\sigma_{11}}{1+r_1}a_1^{1+r_1}-\frac{\sigma_{11}}{1+r_1}\tilde{a}_1^{2\times\frac{1+r_1}{2}}.
	\end{split} 
\end{equation}
Similarly, we also have 
\begin{equation}\nonumber 
	\sigma_{12}\hat{a}_1^{r_2}\tilde{a}_1\leq\frac{2\sigma_{12}}{1+r_2}a_1^{1+r_2}-\frac{\sigma_{12}}{1+r_2}\tilde{a}_1^{2\times\frac{1+r_2}{2}}.
\end{equation}
Thus we have
\begin{equation}\label{eq22}
	\begin{split}
		-\tilde{a}_1\dot{\hat{a}}_1\leq& -\frac{\sigma_{11}}{1+r_1}\tilde{a}_1^{2\times\frac{1+r_1}{2}}-\frac{\sigma_{12}}{1+r_2}\tilde{a}_1^{2\times\frac{1+r_2}{2}}\\&+\frac{2\sigma_{11}}{1+r_1}a_1^{1+r_1}+\frac{2\sigma_{12}}{1+r_2}a_1^{1+r_2}\\&-\rho_1\tilde{a}_1\varphi_1^2\mu_1^2\zeta_1^2.
	\end{split}
\end{equation}
\par From equations \eqref{eq10} and \eqref{eq12}, we have
\begin{equation}\label{eq23}
	\begin{split}
		&\frac{y_2^2}{4}+y_2\dot{y}_2\\&=\frac{y_2^2}{4}+y_2[-\frac{1}{\lambda_2}y_2^{r_1}-\frac{1}{\lambda_2}y_2^{r_2}-\frac{1}{\lambda_2}y_2-\dot{\alpha}_1]\\&=\frac{y_2^2}{4}-\frac{1}{\lambda_2}|y_2|^{1+r_1}-\frac{1}{\lambda_2}|y_2|^{1+r_2}-\frac{1}{\lambda_2}y_2^2-y_2\nu_2\\&\leq\frac{y_2^2}{4}-\frac{1}{\lambda_2}|y_2|^{1+r_1}-\frac{1}{\lambda_2}|y_2|^{1+r_2}-\frac{1}{\lambda_2}y_2^2+\frac{y_2^2}{4}+\nu_2^2\\&=-\frac{1}{\lambda_2}|y_2|^{1+r_1}-\frac{1}{\lambda_2}|y_2|^{1+r_2}-(\frac{1}{\lambda_2}-\frac{1}{2})y_2^2+\nu_2^2\\&\leq-\frac{1}{\lambda_2}(y_2^2)^{\frac{1+r_1}{2}}-\frac{1}{\lambda_2}(y_2^2)^{\frac{1+r_2}{2}}+\nu_2^2
	\end{split}
\end{equation}
for $0<\lambda_2<2$, where $\nu_2=\dot{\alpha}_1=\frac{\partial{\alpha_1}}{\partial{\zeta_1}}\dot{\zeta}_1+\frac{\partial{\alpha_1}}{\partial{\varphi_1}}\dot{\varphi}_1+\frac{\partial{\alpha_1}}{\partial{\psi_1}}\dot{\psi}_1+\frac{\partial{\alpha_1}}{\partial{\hat{a}_1}}\dot{\hat{a}}_1+\frac{\partial{\alpha_1}}{\partial{\mu_1}}\dot{\mu}_1+\frac{\partial{\alpha_1}}{\partial{\dot{\xi}_d}}\ddot{\xi}_d$.
\par Then, applying the inequalities \eqref{eq22} and \eqref{eq23} to \eqref{eq21} yields:
\begin{equation}\label{eq25}
	\begin{split}
		\dot{V}_1\leq&\varphi_1\zeta_1\zeta_2-k_{11}(\zeta_1^2)^{\frac{1+r_1}{2}}-k_{12}(\zeta_1^2)^{\frac{1+r_2}{2}}\\&-\frac{\sigma_{11}}{1+r_1}(\tilde{a}_1^2)^{\frac{1+r_1}{2}}-\frac{\sigma_{12}}{1+r_2}(\tilde{a}_1^2)^{\frac{1+r_2}{2}}\\&-\frac{1}{\lambda_2}(y_2^2)^{\frac{1+r_1}{2}}-\frac{1}{\lambda_2}(y_2^2)^{\frac{1+r_2}{2}}+\Lambda_1,
	\end{split}
\end{equation}
where $\Lambda_1=\frac{2\sigma_{11}}{1+r_1}a_1^{1+r_1}+\frac{2\sigma_{12}}{1+r_2}a_1^{1+r_2}+\nu_2^2+\frac{1}{4\rho_1}$.
\par \textbf{Step i\;(i=2,...,n-1):} Differentiating $\zeta_i$ results in:
\begin{equation}
	\dot{\zeta}_i=\varphi_i[F_i(\bar{x}_{i+1},\xi_{i+1})+\xi_{i+1}]+\psi_i-\dot{\alpha}_{if}.
\end{equation}
Since $\xi_i=\zeta_i+y_i+\alpha_{i-1}$, thus $\dot{\zeta}_i$ can be rewritten as follows:
\begin{equation}
	\dot{\zeta}_i=\varphi_i[F_i(\bar{x}_{i+1},\xi_{i+1})+\zeta_{i+1}+y_{i+1}+\alpha_{i}]+\psi_i-\dot{\alpha}_{if}.
\end{equation}
\par According to the neural network approximation \eqref{eq4}, we approximate $F_i(\bar{x}_{i+1},\xi_{i+1})$ as follows:
\begin{equation}
	F_i(X_i)=W_i^TS(X_i)+\varepsilon_{i}(X_i),
\end{equation}
where $X_i=[\bar{x}_{i+1},\xi_{i+1}]^T$, $W_i\in\mathbb{R}^N$, $S(X_i)\in\mathbb{R}^N$, $\varepsilon_i(X_i)\in\mathbb{R}$. From Assumption \ref{ass4}, it follows that $\vert|W_i\vert|\leq \bar{W}_i$, $|\varepsilon_i|\leq\varepsilon_{i1}$, where $\bar{W}_i$ and $\varepsilon_{i1}$ are unknown positive constants. Define $w_i=\mbox{max}\{\bar{W}_i,\varepsilon_{i1}\}$, then the following inequality holds:
\begin{equation}
	F_i(X_i)\leq w_i\mu_i(X_i),
\end{equation} 
where $\mu_i(X_i)=\vert|S(X_i)\vert|+1$.
\par Design the $i$th virtual controller $\alpha_{i}$ as follows:
\begin{equation}\label{eq29}
	\begin{split}
		\alpha_i=&-\rho_i\hat{a}_i\varphi_i\mu_i^2\zeta_i-\varphi_i\zeta_i-\frac{\psi_i}{\varphi_i}+\frac{\dot{\alpha}_{if}}{\varphi_i}-k_{i1}\zeta_i^{r_1}\\&-k_{i2}\zeta_i^{r_2}-\frac{\varphi_{i-1}}{\varphi_i}\zeta_{i-1},
	\end{split}
\end{equation}
where $k_{i1}$, $k_{i2}$, and $\rho_i$ are positive constants, $r_1$ and $r_2$ are defined in \eqref{eq10}. $\hat{a}_i$ is employed to estimate $a_i=w_i^2$ and determined by the following adaptive law:
\begin{equation}\label{eq31}
	\dot{\hat{a}}_i=\rho_i\varphi_i^2\mu_i^2\zeta_i^2-\sigma_{i1}\hat{a}_i^{r_1}-\sigma_{i2}\hat{a}_i^{r_2},
\end{equation}
where $\sigma_{i1}$ and $\sigma_{i2}$ are positive constants.
\par Using the Young's inequation, the following inequalities can be obtained:
\begin{equation}\label{eq32}
	\begin{aligned}
		\varphi_i\zeta_iF_i&\leq |\varphi_i\zeta_iw_i\mu_i|\leq \rho_iw_i^2\varphi_i^2\mu_i^2\zeta_i^2+\frac{1}{4\rho_i},\\
		\varphi_i\zeta_iy_{i+1}&\leq \varphi_i^2\zeta_i^2+\frac{y_{i+1}^2}{4},
	\end{aligned}
\end{equation}
where $\rho_i$ is defined in \eqref{eq29}.
\par Next, we construct the Lyapunov function $V_i$ as follows:
\begin{equation}
	V_i=\frac{1}{2}\zeta_i^2+\frac{1}{2}\tilde{a}_i^2+\frac{1}{2}y_{i+1}^2,
\end{equation}
where $\tilde{a}_i=a_i-\hat{a}_i$.
\par Using the virtual controller $\alpha_{i}$ \eqref{eq29}, the adaptive law \eqref{eq31}, and inequalities \eqref{eq32}, and calculating the derivative of $V_i$, we have
\begin{equation}\label{eq35}
	\begin{split}
		\dot{V}_i\leq& \rho_i\tilde{a}_i\varphi_i^2\mu_i^2\zeta_i^2+ \varphi_i\zeta_i\zeta_{i+1}-\varphi_{i-1}\zeta_{i-1}\zeta_{i}-k_{i1}\zeta_i^{1+r_1}\\&-k_{i2}\zeta_i^{1+r_2}+\frac{y_{i+1}^2}{4}+\frac{1}{4\rho_i}+\tilde{a}_i\dot{\tilde{a}}_i+y_{i+1}\dot{y}_{i+1}\\\leq&\varphi_i\zeta_i\zeta_{i+1}-\varphi_{i-1}\zeta_{i-1}\zeta_{i}-k_{i1}\zeta_i^{1+r_1}-k_{i2}\zeta_i^{1+r_2}\\&+\frac{y_{i+1}^2}{4}+\frac{1}{4\rho_i}
		+\sigma_{i1}\tilde{a}_i\hat{a}_i^{r_1}+\sigma_{i2}\tilde{a}_i\hat{a}_1^{r_2}+y_{i+1}\dot{y}_{i+1}.
	\end{split}
\end{equation}
\par For the adaptive law \eqref{eq31}, it follows that $\hat{a}_i(t)>0$, $\forall t>0$ for a positive initial value $\hat{a}_i(0)$. Thus $a_i>\tilde{a}_i$ when $\hat{a}_i(0)>0$. According to Lemma \ref{lem2} and Lemma \ref{lem3}, we have
\begin{equation}\label{eq36}
	\begin{split}
		\sigma_{i1}\hat{a}_i^{r_1}\tilde{a}_i&=\sigma_{i1}\hat{a}_i^{r_1}(a_i-\hat{a}_i)\\&\leq\sigma_{i1}\frac{1}{1+r_1}(a_i^{1+r_1}-\hat{a}_i^{1+r_1})\\&=\sigma_{i1}\frac{1}{1+r_1}(a_i^{1+r_1}-[a_i-\tilde{a}_i]^{1+r_1})\\&\leq\sigma_{i1}\frac{1}{1+r_1}(a_i^{1+r_1}+a_i^{1+r_1}-\tilde{a}_i^{1+r_1})\\&=\frac{2\sigma_{i1}}{1+r_1}a_i^{1+r_1}-\frac{\sigma_{i1}}{1+r_1}\tilde{a}_i^{2\times\frac{1+r_1}{2}}.
	\end{split}
\end{equation}
\par Similarly, we also have 
\begin{equation}\label{eq37}
	\sigma_{i2}\hat{a}_i^{r_2}\tilde{a}_i\leq \frac{2\sigma_{i2}}{1+r_2}a_i^{1+r_2}-\frac{\sigma_{i2}}{1+r_2}\tilde{a}_i^{2\times\frac{1+r_2}{2}}.
\end{equation}
\par Furthermore, from the definitions of $y_i$ in \eqref{eq12} and the dynamic variable $\alpha_{if}$ in \eqref{eq10},  we have
\begin{equation}\nonumber
	\dot{y}_i=-\frac{1}{\lambda_i}y_i^{r_1}-\frac{1}{\lambda_i}y_i^{r_2}-\frac{1}{\lambda_i}y_i-\nu_i,
\end{equation}
where $\nu_i=\dot{\alpha}_{i-1}$. Then we have
\begin{equation}\label{eq38}
	\begin{split}
		&\frac{y_{i+1}^2}{4}+y_{i+1}\dot{y}_{i+1}\\&=-\frac{1}{\lambda_{i+1}}\bigg(|y_{i+1}|^{1+r_1}+|y_{i+1}|^{1+r_2}+y_{i+1}^2\bigg)\\&\quad+\frac{y_{i+1}^2}{4}-y_{i+1}\nu_{i+1}\\&\leq-\frac{1}{\lambda_{i+1}}|y_{i+1}|^{1+r_1}-\frac{1}{\lambda_{i+1}}|y_{i+1}|^{1+r_2}\\&\quad-(\frac{1}{\lambda_{i+1}}-\frac{1}{2})y_{i+1}^2+\nu_{i+1}^2\\&\leq-\frac{1}{\lambda_{i+1}}(y_{i+1}^2)^{\frac{1+r_1}{2}}-\frac{1}{\lambda_{i+1}}(y_{i+1}^2)^{\frac{1+r_2}{2}}+\nu_{i+1}^2
	\end{split}
\end{equation}
when $0<\lambda_{i+1}<2$.
\par Applying the inequalities\eqref{eq36}-\eqref{eq38} to \eqref{eq35}, we have
\begin{equation}\label{eq39}
	\begin{split}
		\dot{V}_i\leq&\varphi_i\zeta_i\zeta_{i+1}-\varphi_{i-1}\zeta_{i-1}\zeta_{i}-k_{i1}(\zeta_i^2)^{\frac{1+r_1}{2}}-k_{i2}(\zeta_i^2)^{\frac{1+r_2}{2}}\\&-\frac{\sigma_{i1}}{1+r_1}(\tilde{a}_i^2)^{\frac{1+r_1}{2}}-\frac{\sigma_{i2}}{1+r_2}(\tilde{a}_i^2)^{\frac{1+r_2}{2}}\\&-\frac{1}{\lambda_{i+1}}(y_{i+1}^2)^{\frac{1+r_1}{2}}-\frac{1}{\lambda_{i+1}}(y_{i+1}^2)^{\frac{1+r_2}{2}}+\Lambda_i,
	\end{split}
\end{equation}
where $\Lambda_i=\frac{2\sigma_{i1}}{1+r_1}a_i^{1+r_1}+\frac{2\sigma_{i2}}{1+r_2}a_i^{1+r_2}+\nu_{i+1}^2+\frac{1}{4\rho_i}$.
\par \textbf{Step n:} In this step, a fixed-time safe controller will be given. Differentiating $\zeta_n$ yields:
\begin{equation}
	\dot{\zeta}_n=\varphi_n[F_n(\bar{x}_n)+g_nu]+\psi_n-\dot{\alpha}_{nf}.
\end{equation}
\par It is noted that $F_n(\bar{x}_n)$ can be approximated by:
\begin{equation}\nonumber
	F_n(\bar{x}_n)=W_n^TS(\bar{x}_n)+\varepsilon_n(\bar{x}_n)\leq w_n\mu_n(\bar{x}_n),
\end{equation}
where $\vert|W_n\vert|\leq\bar{W}_n$, $|\varepsilon_n|\leq\varepsilon_{n1}$, $w_n=\mbox{max}\{\bar{W}_n,\varepsilon_{n1}\}$, and $\bar{W}_n$, $\varepsilon_{n1}$ are unknown positive constants. According to Young's inequality, we have the following inequalities:
\begin{equation}\nonumber
	\varphi_n\zeta_nF_n\leq \underline{g}_n\rho_n\varphi_n^2w_n^2\mu_n^2\zeta_n^2+\frac{1}{4\rho_n \underline{g}_n},
\end{equation}
\begin{equation}\nonumber
	\psi_n\zeta_n\leq \underline{g}_n\psi_n^2\zeta_n^2+\frac{1}{4\underline{g}_n},
\end{equation}
\begin{equation}\nonumber
	-\zeta_n\dot{\alpha}_{nf}\leq \underline{g}_n\dot{\alpha}_{nf}^2\zeta_n^2+\frac{1}{4\underline{g}_n},
\end{equation}
where $\rho_n>0$,  $\underline{g}_n>0$ is the lower bound of $g_n$. Define $a_n=w_n^2$, then we have 
\begin{equation}\nonumber
	\begin{split}
		\zeta_n\dot{\zeta}_n&=\varphi_n\zeta_nF_n+\varphi_n\zeta_ng_nu+\zeta_n\psi_n-\zeta_n\dot{\alpha}_{nf}\\&\leq\underline{g}_n\rho_n\varphi_n^2 a_n\mu_n^2\zeta_n^2+\frac{1}{4\rho_n\underline{g}_n}+\underline{g}_n\psi_n^2\zeta_n^2+\underline{g}_n\dot{\alpha}_{nf}^2\zeta_n^2\\&\quad+\frac{1}{2\underline{g}_n}+\varphi_n\zeta_n g_nu.
	\end{split}
\end{equation}

\par It is time to design the controller $u$ as follows:
\begin{equation}\label{eq41}
	\begin{split}
		u=&-\rho_n\varphi_n\hat{a}_n\mu_n^2\zeta_n-\frac{\psi_n^2\zeta_n}{\varphi_n}-\frac{\dot{\alpha}_{nf}^2\zeta_n}{\varphi_n}\\&-\frac{\varphi_{n-1}^2\zeta_{n-1}^2\zeta_n}{\varphi_n}-\frac{k_{n1}}{\varphi_n}\zeta_n^{r_1}-\frac{k_{n2}}{\varphi_n}\zeta_n^{r_2},
	\end{split}
\end{equation}
where $\hat{a}_n$ is the estimation of $a_n$ and determined by the following adaptive law:
\begin{equation}\label{eq422}
	\dot{\hat{a}}_n=\rho_n\varphi_n^2\mu_n^2\zeta_n^2-\sigma_{n1}\hat{a}_n^{r_1}-\sigma_{n2}\hat{a}_n^{r_2},
\end{equation}
and $\sigma_{n1}$, $\sigma_{n2}$ are positive constants.
\par When we choose  $\hat{a}_n(0)\geq0$, we have $\hat{a}_n\geq0$. Then, 
according to Assumption \ref{ass2}, we have
\begin{equation}\nonumber
	\begin{split}
		&\varphi_n\zeta_ng_nu\\&=-g_n\rho_n\varphi_n^2\hat{a}_n\mu_n^2\zeta_n^2-g_n\psi_n^2\zeta_n^2-g_n\dot{\alpha}_{nf}^2\zeta_{n}^2\\&\quad-g_n\varphi_{n-1}^2\zeta_{n-1}^2\zeta_n^2-g_nk_{n1}\zeta_n^{1+r_1}-g_nk_{n2}\zeta_n^{1+r_2}\\&\leq-\underline{g}_n\rho_n\varphi_n^2\hat{a}_n\mu_n^2\zeta_n^2-\underline{g}_n\psi_n^2\zeta_n^2-\underline{g}_n\dot{\alpha}_{nf}^2\zeta_n^2\\&\quad-\underline{g}_n\varphi_{n-1}^2\zeta_{n-1}^2\zeta_n^2-\underline{g}_nk_{n1}\zeta_n^{1+r_1}-\underline{g}_nk_{n2}\zeta_n^{1+r_2}.
	\end{split}
\end{equation}
\par Substituting the above inequality to  $\zeta_n\dot{\zeta}_n$ yields:
\begin{equation}\label{eq42}
	\begin{split}
		\zeta_n\dot{\zeta}_n&\leq \underline{g}_n\rho_n\tilde{a}_n\varphi_n^2\mu_n^2\zeta_n^2-\underline{g}_nk_{n1}\zeta_n^{1+r_1}-\underline{g}_nk_{n2}\zeta_n^{1+r_2}\\&-\underline{g}_n\varphi_{n-1}^2\zeta_{n-1}^2\zeta_n^2+\frac{1}{4\rho_n\underline{g}_n}+\frac{1}{2\underline{g}_n}.
	\end{split}
\end{equation}
\par Now we construct a Lyapunov function $V_n$ as follows:
\begin{equation}
	V_n=\frac{1}{2}\zeta_n^2+\frac{1}{2}\tilde{a}_n^2. 
\end{equation}
Using the inequality \eqref{eq42} and the adaptive law \eqref{eq422}, we have
\begin{equation}\label{eq45}
	\begin{split}
		\dot{V}_n\leq& -\underline{g}_nk_{n1}\zeta_n^{1+r_1}-\underline{g}_nk_{n2}\zeta_n^{1+r_2}-\underline{g}_n\varphi_{n-1}^2\zeta_{n-1}^2\zeta_n^2\\&+\frac{1}{4\rho_n\underline{g}_n}+\frac{1}{2\underline{g}_n}-\underline{g}_n\sigma_{n1}\tilde{a}_n\hat{a}_n^{r_1}-\underline{g}_n\sigma_{n2}\tilde{a}_n\hat{a}_n^{r_2}\\\leq&-\underline{g}_nk_{n1}(\zeta_n^2)^{\frac{1+r_1}{2}}-\underline{g}_nk_{n2}(\zeta_n^2)^{\frac{1+r_2}{2}}-\underline{g}_n\varphi_{n-1}^2\zeta_{n-1}^2\zeta_n^2\\&-\underline{g}_n\frac{\sigma_{n1}}{1+r_1}(\tilde{a}_n^2)^{\frac{1+r_1}{2}}-\underline{g}_n\frac{\sigma_{n2}}{1+r_2}(\tilde{a}_n^2)^{\frac{1+r_2}{2}}+\Lambda_n,
	\end{split}
\end{equation}
where $\Lambda_n=\underline{g}_n\frac{2\sigma_{n1}}{1+r_1}a_n^{1+r_1}+\underline{g}_n\frac{2\sigma_{n2}}{1+r_2}a_n^{1+r_2}+\frac{1}{4\rho_n\underline{g}_n}+\frac{1}{2\underline{g}_n}$.

Define $$V=V_1+...,V_n.$$ Using the inequalities \eqref{eq25}, \eqref{eq39}, and \eqref{eq45}, we have
\begin{equation}\nonumber
	\begin{split}
		\dot{V}&=\sum_{i=1}^n\dot{V}_i\\&\leq-\sum_{i=1}^{n-1}k_{i1}(\zeta_i^2)^{\frac{1+r_1}{2}}-\sum_{i=1}^{n-1}k_{i2}(\zeta_i^2)^{\frac{1+r_2}{2}}\\&\quad-\sum_{i=1}^{n-1}\frac{\sigma_{i1}}{1+r_1}(\tilde{a}_i^2)^{\frac{1+r_1}{2}}-\sum_{i=1}^{n-1}\frac{\sigma_{i2}}{1+r_2}(\tilde{a}_i^2)^{\frac{1+r_2}{2}}\\&\quad-\sum_{i=2}^{n}\frac{1}{\lambda_i}(y_i^2)^{\frac{1+r_1}{2}}-\sum_{i=2}^{n}\frac{1}{\lambda_i}(y_i^2)^{\frac{1+r_2}{2}}-\underline{g}_nk_{n1}(\zeta_n^2)^{\frac{1+r_1}{2}}\\&\quad-\underline{g}_nk_{n2}(\zeta_n^2)^{\frac{1+r_2}{2}}-\frac{\underline{g}_n\sigma_{n1}}{1+r_1}(\tilde{a}_n^2)^{\frac{1+r_1}{2}}-\frac{\underline{g}_n\sigma_{n2}}{1+r_2}(\tilde{a}_n^2)^{\frac{1+r_2}{2}}\\&\quad-\underline{g}_n\varphi_{n-1}^2\zeta_{n-1}^2\zeta_n^2+\varphi_{n-1}\zeta_{n-1}\zeta_{n}+\sum_{i=1}^n\Lambda_i.
	\end{split}
\end{equation}
\par It is noted that 
\begin{equation}\nonumber
	\varphi_{n-1}\zeta_{n-1}\zeta_n\leq\underline{g}_n\varphi_{n-1}^2\zeta_{n-1}^2
	\zeta_n^2+\frac{1}{4\underline{g}_n},
\end{equation}
then we have
\begin{equation}\label{eq46}
	\begin{split}
		\dot{V}&\leq-\sum_{i=1}^{n-1}k_{i1}(\zeta_i^2)^{\frac{1+r_1}{2}}-\sum_{i=1}^{n-1}k_{i2}(\zeta_i^2)^{\frac{1+r_2}{2}}\\&\quad-\sum_{i=1}^{n-1}\frac{\sigma_{i1}}{1+r_1}(\tilde{a}_i^2)^{\frac{1+r_1}{2}}-\sum_{i=1}^{n-1}\frac{\sigma_{i2}}{1+r_2}(\tilde{a}_i^2)^{\frac{1+r_2}{2}}\\&\quad-\sum_{i=2}^{n}\frac{1}{\lambda_i}(y_i^2)^{\frac{1+r_1}{2}}-\sum_{i=2}^{n}\frac{1}{\lambda_i}(y_i^2)^{\frac{1+r_2}{2}}-\underline{g}_nk_{n1}(\zeta_n^2)^{\frac{1+r_1}{2}}\\&\quad-\underline{g}_nk_{n2}(\zeta_n^2)^{\frac{1+r_2}{2}}-\frac{\underline{g}_n\sigma_{n1}}{1+r_1}(\tilde{a}_n^2)^{\frac{1+r_1}{2}}-\frac{\underline{g}_n\sigma_{n2}}{1+r_2}(\tilde{a}_n^2)^{\frac{1+r_2}{2}}\\&\quad+\Lambda,
	\end{split}
\end{equation}
where $\Lambda=\frac{1}{4\underline{g}_n}+\sum_{i=1}^n\Lambda_i$. 

\section{Proof of Theorem 1}\label{appendix2}

Proof. 
Define
\begin{equation}
\Xi_1=2^{\frac{1+r_1}{2}}\min\limits_{i=1,...,n-1}\{k_{i1},\underline{g}_nk_{n1},\frac{\sigma_{i1}}{1+r_1},\frac{\underline{g}_n\sigma_{n1}}{1+r_1},\frac{1}{\lambda_i}\},
\end{equation}
\begin{equation}
	\Xi_2=2^{\frac{1+r_2}{2}}\min\limits_{i=1,...,n-1}\{k_{i2},\underline{g}_nk_{n2},\frac{\sigma_{i2}}{1+r_2},\frac{\underline{g}_n\sigma_{n2}}{1+r_2},\frac{1}{\lambda_i}\},
\end{equation}
where parameters $k_{i1}$, $k_{i2}$ are gains given in virtual controllers and should be positive. $g_n$, $\sigma_{i1}$, $\sigma_{i2}$, and $\lambda_i$ are  positive constants. $r_1=\frac{m}{n}$, $r_2=\frac{p}{q}$, and $m<n$, $p>q$ are positive odd constants.

 From the inequality \eqref{eq46}, we have
\begin{equation}\nonumber
	\begin{split}
		\dot{V}\leq&-\Xi_1\sum_{i=1}^{n-1}\bigg[(\frac{1}{2}\zeta_i^2)^{\frac{1+r_1}{2}}+(\frac{1}{2}\tilde{a}_i^2)^{\frac{1+r_1}{2}}+(\frac{1}{2}y_{i+1}^2)^{\frac{1+r_1}{2}}\bigg]\\&-\Xi_1\bigg[(\frac{1}{2}\zeta_n^2)^{\frac{1+r_1}{2}}+(\frac{1}{2}\tilde{a}_n^2)^{\frac{1+r_1}{2}}\bigg]\\&-\Xi_2\sum_{i=1}^{n-1}\bigg[(\frac{1}{2}\zeta_i^2)^{\frac{1+r_2}{2}}+(\frac{1}{2}\tilde{a}_i^2)^{\frac{1+r_2}{2}}+(\frac{1}{2}y_{i+1}^2)^{\frac{1+r_2}{2}}\bigg]\\&-\Xi_2\bigg[(\frac{1}{2}\zeta_n^2)^{\frac{1+r_2}{2}}+(\frac{1}{2}\tilde{a}_n^2)^{\frac{1+r_2}{2}}\bigg]+\Lambda. 
	\end{split}
\end{equation}
\par Furthermore, according to Lemma \ref{lem4}, we have:
\begin{equation}
	\begin{split}
		\dot{V}\leq&-\Xi_1\sum_{i=1}^{n}\bigg(V_i^{\frac{1+r_1}{2}}\bigg)-\Xi_2\sum_{i=1}^{n-1}\bigg(3^{\frac{1-r_2}{2}}V_i^{\frac{1+r_2}{2}}\bigg)\\&-\Xi_2\bigg(2^{\frac{1-r_2}{2}}V_n^{\frac{1+r_2}{2}}\bigg)+\Lambda\\\leq&-\Xi_1V^{\frac{1+r_1}{2}}-3^{\frac{1-r_2}{2}}\Xi_2\bigg(n^{\frac{1-r_2}{2}}V^{\frac{1+r_2}{2}}\bigg)+\Lambda\\\leq&-\Xi_1V^{\frac{1+r_1}{2}}-\bar{\Xi}_2V^{\frac{1+r_2}{2}}+\Lambda,
	\end{split}
\end{equation}
where $\bar{\Xi}_2=(3n)^{\frac{1-r_2}{2}}\Xi_2$.
\par Finally, from Lemma \ref{lem1},  there exists a settling time $T$ such that
\begin{equation*}
	V\leq \mathcal{R}= \min\Big\{\big(\frac{\Lambda}{\Xi_1\theta}\big)^{\frac{2}{1+r_1}},\big(\frac{\Lambda}{\bar{\Xi}_2\theta}\big)^{\frac{2}{1+r_2}}\Big\},
\end{equation*}
when $t\geq T$, where 	
\begin{equation*}
	T\leq\frac{2}{\Xi_1(1-\theta)(1-r_1)}+\frac{2}{\bar{\Xi}_2(1-\theta)(r_2-1)},
\end{equation*}
and $0<\theta<1$ is a constant. 
\par Therefore, the closed-loop system \eqref{eq9} is practically fixed-time stable. Since $V\in\mathcal{L}_{\infty}$, then all the states of the closed-loop system are bounded, that is, $\zeta_i\in\mathcal{L}_\infty$, $\tilde{a}_i\in\mathcal{L}_\infty$, and $y_{i}\in\mathcal{L}_\infty$.
\par Furthermore, it is noted that $V\leq\mathcal{R}$ when $t\geq T$, thus we have $|\zeta_1|\leq\sqrt{2\mathcal{R}}$ when $t\geq T$. Recalling the transformation functions \eqref{eq6} and \eqref{eq7} and applying the mean value theorem,  there exists a constant $\hat{\xi}$ such that
\begin{equation}\nonumber
	\begin{split}
		|x_1-y_d|=&\bigg|\frac{h_{11}(t)+h_{12}(t)}{2}\tanh\bigg(\frac{2\xi_1}{h_{11}(t)+h_{12}(t)}\bigg)\\&-\frac{h_{11}(t)+h_{12}(t)}{2}\tanh\bigg(\frac{2\xi_d}{h_{11}(t)+h_{12}(t)}\bigg)\bigg|\\=&\frac{h_{11}(t)+h_{12}(t)}{2}\bigg|\frac{1}{cosh^2(\hat{\xi})}\times\frac{2(\xi_1-\xi_d)}{h_{11}(t)+h_{12}(t)}\bigg|\\\leq& \frac{h_{11}(t)+h_{12}(t)}{2}\times \frac{2}{h_{11}(t)+h_{12}(t)}\big|\xi_1-\xi_d\big|\\=&|\xi_1-\xi_d|\\=&\zeta_1,
	\end{split}
\end{equation}
where $\hat{\xi}\in(\frac{2\xi_1}{h_{11}(t)+h_{12}(t)},\frac{2\xi_d}{h_{11}(t)+h_{12}(t)})$. 
Thus the output tracking error satisfies  $|e|=|x_1-y_d|\leq|\zeta_1|\leq\sqrt{2\mathcal{R}}$ when $t\geq T$, which means that the tracking error is bounded in fixed time.
\par Finally, we verify that all the state constraints are  satisfied all the time. According to   Assumption \ref{ass3}, we know that $\xi_d$ is bounded. It is noted that $\zeta_i\in\mathcal{L}_\infty$, thus $\zeta_1$ is also bounded. Therefore, $\xi_1=\zeta_1+\xi_d$ is bounded, which implies that $-h_{11}(t)<x_1(t)<h_{12}(t)$ is satisfied when $-h_{11}(0)<x_1(0)<h_{12}(0)$.  Additionally, according to the fact that $y_2\in\mathcal{L}_\infty$,   $\alpha_{2f}=y_2+\alpha_1$ is bounded, and $\dot{\alpha}_{2f}$ is bounded. Then, the boundedness of $\zeta_2$ and $\alpha_{2f}$ ensures that $\xi_2=\zeta_2+\alpha_{2f}$ is bounded, which implies that the constraint $-h_{21}(t)<x_2(t)<h_{22}(t)$ is satisfied when $-h_{21}(0)<x_2(0)<h_{22}(0)$. Using similar analysis, it can be shown that all the state constraints are satisfied all the time. 
\par The proof is thus complete.

\makecontacts

\end{document}